\documentclass[a4paper]{jpconf}
\usepackage{graphicx}
\def\beqar {\begin{eqnarray}}
\def\eeqar {\end{eqnarray}}
\def\beq {\begin{equation}}
\def\eeq {\end{equation}}
\def\ra {{\rangle}}
\def\la {{\langle}}

\def\Tr {{\rm Tr}}
\def\tr {{\rm tr}}

\def\del {{\partial}}
\def\bdel{\bar{\partial}}

\def\l{{\lambda}}

\def\bi {{\bar i}}

\def \A {{\cal A}}
\def \D {{\cal D}}

\def \L {{\cal L}}

\def\no2 {{\textstyle{n\over 2}}}

\begin{document}
\title{Quantum Hall Effect, Bosonization and Chiral Actions in Higher Dimensions}

\author{Dimitra Karabali}

\address{Department of Physics and Astronomy, Lehman College, CUNY, Bronx, NY 10468, USA}

\ead{dimitra.karabali@lehman.cuny.edu}

\begin{abstract}
We give a brief review of the Quantum Hall effect in higher dimensions and its relation to fuzzy spaces. For a quantum Hall system, the lowest Landau level dynamics is given by a one-dimensional matrix action. This can be used to write down a bosonized noncommutative field theory describing the interactions of higher dimensional nonrelativistic fermions with abelian or nonabelian gauge fields in the lowest Landau level. 
This general approach is applied explicitly to the case of QHE on ${\bf CP}^k$. It is shown that in the semiclassical limit the effective action  contains a bulk Chern-Simons type term whose anomaly is exactly canceled by a boundary term given in terms of a chiral, gauged Wess-Zumino-Witten action suitably generalized to higher dimensions.
\end{abstract}

\section{Introduction}
Quantum Hall effect in two dimensions is a very important physical phenomenon \cite{2dqhe}. In addition to a variety of fascinating experimental results it has provided a clear theoretical framework for exploring a number of field/string theory ideas such as bosonization, conformal invariance, topological field theories, noncommutative geometry, $D$-brane physics, etc. The study of QHE in more general contexts, such as higher dimensions and different geometries, has recently attracted a lot of attention \cite{HZ}-\cite{everyone}, following the original work of Zhang and Hu who studied the QHE on $S^4$ with an $SU(2)$ background magnetic field \cite{HZ}. 

In this brief review I will outline our work on the formulation of the QHE on arbitrary even dimensions with particular focus on the edge and bulk dynamics of the corresponding higher dimensional quantum Hall droplets in the presence of external gauge fields and the connection to noncommutative field theories \cite{KN1}-\cite{nair}. Applications of these results beyond the QHE framework are mentioned at the end.

\section{Matrix formulation of quantum Hall dynamics}

The basic phenomenon of QHE refers to the dynamics of charged fermions on a manifold in the presence of a uniform magnetic field. At the single particle level, the energy eigenstates are grouped into the Landau levels. For high values of the magnetic field at low temperatures, the separation of levels is high compared to the available thermal excitation energy and the dynamics is confined to the lowest Landau level. Although our detailed formulation of higher dimensional QHE  is based on the Landau problem on ${\bf CP}^k$, we start with a general matrix formulation of the dynamics of noninteracting fermions in the lowest Landau level, which eventually leads to a bosonization approach in terms of a noncommutative field theory. 

Let $N$ denote the dimension of the one-particle Hilbert space corresponding to the states of the lowest Landau level, $K$ of which are occupied by fermions. The spin degree of freedom is neglected, so each state can be occupied by a single fermion. In a physical sample, there is also a potential $\hat{V}$ which confines the fermions to within the sample. The fermions are localized around the minimum of the potential but, because of the exclusion principle, they spread over an area forming an incompressible droplet. The droplet is mathematically characterized by a diagonal density matrix 
 $\hat {\rho}_0$ which is equal to 1 for occupied states and zero for unoccupied states. The most general fluctuations which preserve the LLL condition and the number of occupied states are unitary transformations of $\hat {\rho}_0$, namely $\hat {\rho}_0 \rightarrow \hat{\rho}=\hat{U}  \hat {\rho}_0 \hat{U} ^ \dagger$, where  $\hat{U}$ is an $(N \times N)$ unitary matrix. The
action which determines ${\hat U}$ is given by 
\beq
S_{0}=  \int dt~ \Tr \left[ i  {\hat \rho}_0 { \hat U}^\dagger \del_t {\hat U}
~-~ {\hat \rho}_0 {\hat U}^\dagger {\hat{V}} {\hat U} \right]
\label{1}
\eeq
(The Hamiltonian is $\hat{V}$ up to an additive constant.) $\hat{U}$ can be thought of as a collective variable describing all the possible excitations within the LLL. The equation of motion resulting from (\ref{1}) is the expected evolution equation for the density matrix $\hat{\rho}$, namely
\beq
i {{\partial \hat{\rho}} \over {\partial t}} = [ \hat{V} , \hat{\rho}]
\label{3}
\eeq

The action $S_{0}$ can also be written as 
\beq
S_{0}= {N \over N'}  \int d\mu dt~ \tr~\left[ i ({\rho}_0 *{  U}^\dagger  * \del_t { U})
~-~ ({ \rho}_0 *{U}^\dagger  * {{V}} * {U}) \right]
\label{4a}
\eeq
where $d\mu$ is the volume measure of the space where QHE has been defined and $\rho_0,~U,~V$ are the symbols of the corresponding matrices on this space. (The hatted expressions correspond to matrices and unhatted ones to the corresponding symbols, which are fields on the space where QHE is defined.) Equation (\ref{4a}) is written for the case of nonabelian fermions coupled to a background gauge field in some representation $J'$ of dimension $N'$; the corresponding symbols are $(N' \times N')$ matrix valued functions. 
Our notation is such that ``$\Tr$" indicates trace over the $N$-dimensional LLL Hilbert space while ``$\tr$" indicates trace over the $N'$-dimensional representation $J'$. In the case of abelian fermions, $N'=1$ and $\tr$ is trivial. 

The general definitions of the symbol and star-product are as follows. If $\Psi_m(\vec{x})$, $m=1,\cdots,N$,  represent the correctly normalized LLL single-particle wavefunctions, then the definition of the symbol corresponding to a $(N \times N)$ matrix $\hat{O}$, with matrix elements $O_{ml}$ is
\beq
O(\vec{x}, t) = { 1 \over N} \sum_{m,l} \Psi_m(\vec{x}) O_{ml}(t) \Psi^*_l(\vec{x})
\label{symb}
\eeq
The star-product is defined as 
\beq
\big(\hat{O}_1\hat{O}_2\big)_{symbol} = O_1(\vec{x}, t) * O_2 (\vec{x}, t) 
\label{star}
\eeq
The action $S_{0}$ in (\ref{1}), or equivalently (\ref{4a}), provides an exact bosonization for the original noninteracting fermion problem. It does not explicitly depend on the particular space and its dimensionality or the abelian or nonabelian nature of the underlying fermionic system. This information is encoded in equation (\ref{4a}) in the definition of the symbol, the star product and the measure.

This matrix formulation can be extended to include external, fluctuating gauge fields \cite{KA, review} (in addition to the uniform background magnetic field which defines the Landau problem).  Gauge interactions are described by a matrix action $S$ which is the gauged version of $S_0$,
\beq
S= \int dt~ \Tr \left[ i  \hat {\rho}_0  \hat {U}^\dagger ( \del_t + i \hat{\A} ) \hat {U}
~-~  \hat {\rho}_0 \hat {U}^\dagger \hat{V} \hat{U} \right]
\label{6}
\eeq
where $\hat{\A}$ is a matrix gauge potential. It can also be written in terms of the corresponding symbols as
 \beq
S= {N \over N'}   \int dt~d\mu ~ \tr~\left[ i  \rho_0 * U^\dagger  * \del_t U
~-~ \rho_0 *U^\dagger * V *U- \rho_0 *U^\dagger * \A *U \right]
\label{10}
\eeq
The action (\ref{10}) is now invariant under the infinitesimal transformations
\beqar
\delta U & =& - i \l * U \nonumber \\
\delta \A (\vec{x}, t)  & = & \del _t \l (\vec{x}, t) - i \left( \l* (V+\A) - (V + \A) * \l \right) 
\label{11}
\eeqar

The key physical question is how the field $\A$ is related to the gauge fields $A_{\mu}$ to which the original fermions couple. Once this is known, the action (\ref{10}) can be expressed in terms of the usual gauge fields. Since $S$ describes gauge interactions it has to be invariant under the usual gauge transformation
 \beqar
 \delta A_{\mu} & = & \del_{\mu} \Lambda + i [\bar{A}_\mu + A_\mu , ~ \Lambda] 
 \label{13} \\
 \delta \bar{A}_{\mu} & = & 0 \nonumber 
 \eeqar
 Here $\Lambda$ is the infinitesimal gauge parameter and $\bar{A}_\mu$ is a possible nonabelian background field. What we need is an expression for $\A$ in terms of $A_{\mu}$ such that when the gauge fields are transformed as in (\ref{13}), the field $\A$ undergoes the transformation (\ref{11}). In other words, the transformation (\ref{13}) is induced by the transformation (\ref{11}). Since ${\A}$ is the time-component of a noncommutative gauge field, the relation between ${\A}$ and the commutative gauge fields $A_\mu$ is essentially a Seiberg-Witten transformation \cite{seib1,seib2}. The bosonized action of the LLL fermionic system in the presence of gauge interactions then follows in a straightforward way \cite{KA, review}.

This approach, which is based on a matrix formulation, provides a very general way to construct the bosonic action describing the dynamics of  the underlying LLL fermionic system in any space that admits a consistent formulation of QHE.  This is a generalization of a method used by Sakita \cite{sakita2} to derive the electromagnetic interactions of LLL spinless electrons in the two-dimensional plane. The action $S_0$ in (\ref{1}) was also used in the context of one-dimensional free fermions and their relation to $c=1$ string theory \cite{wadia}. Below we shall show how these general ideas get implemented in the case of $\nu=1$ QHE on ${\bf CP}^k$, where $\nu$ is the filling fraction. We find, in the limit where $N \rightarrow \infty$ and the number of fermions is large, that the action $S$ separates into a boundary term describing the coupling of the quantum Hall droplet to the external gauge field $A_\mu$, and a purely $A_\mu$-dependent bulk term, which is a Chern-Simons like term. An anomaly cancellation between these two terms renders the action gauge invariant, as expected.

\section{Quantum Hall effect on  ${\bf CP}^k$}

Our analysis of the Landau problem on ${\bf CP}^k$ is based on the fact that this is a coset manifold, ${\bf CP}^k = SU(k+1) / U(k)$ \cite{KN1}-\cite{KN3}. Here we shall briefly describe the group theoretical method we have used to obtain the LLL wavefunctions entering the definition of the symbol and the star-product in (\ref{symb}, \ref{star}), necessary ingredients in deriving the action (\ref{10}). Explicit details on the full Landau spectrum on ${\bf CP}^k$ are given in \cite{KN3}.

${\bf CP}^k$ can be parametrized in terms of a $(k+1) \times (k+1)$ matrix $g$ in the fundamental representation of $SU(k+1)$, by making the identification $g \sim g h$, where $h~\in~U(k)$. For ${\bf CP}^k$ one can have both abelian and nonabelian uniform magnetic fields. (The corresponding field strengths are proportional to the Riemannian curvature which is constant, proportional to the $U(k)$ structure constants, in the basis of the frame fields.) The $U(1)$ and $SU(k)$ background gauge fields are given by
\beqar
 a  &=&  i n \sqrt{{{2k} \over {k+1}}} \tr (t_{k^2+2k} g^{-1} dg ) ~~,~~~~~~~~~~~da=n\Omega \nonumber \\
\bar{A}^a & =&  2i \tr (t^a g^{-1} dg) \label{NA}
\eeqar
The $U(1)$ field strength is proportional to the K\"ahler two-form $\Omega$. The constant $n$ can be expressed in terms of the radius $R$ of ${\bf CP}^k$ and the $U(1)$ magnetic field $B$ as, $n=2B R^2$. Notice that $\bar{A}$ does not depend on $n$.
$t_A$ are the generators of $SU(k+1)$ as matrices in the fundamental representation, normalized so that $\tr (t_A t_B) = {1 \over 2} \delta_{AB}$. They are classified into three groups: $t_a$, $a =1, ~2, \cdots , ~ k^2 -1$, corresponding to the $SU(k)$ part of
$U(k) \subset SU(k+1)$; $t_{k^2+2k}$, the $U(1)$
direction of the subgroup $U(k)$ and the remaining coset generators $t_\alpha$, $\alpha = 1,\cdots, 2k$. The latter ones can be further separated into the raising and lowering type $t_{\pm I}= t_{2I-1} \pm i t_{2I}, ~ I = 1, \cdots,k$. 

The Landau problem on ${\bf CP}^k$ is defined by the choice of ``constant" magnetic fields (\ref{NA}). The corresponding wavefunctions form an $SU(k+1)$ representation and can be uniformly expressed in terms of the Wigner ${\D}$-functions which are the matrices corresponding to the group elements $g$ in a particular representation $J$. Taking into account the proper normalization we have,
\beq
\Psi= \sqrt{N} {\D}^{(J)} _{L,R} (g) = \sqrt{N} ~\la J, l_A \vert~{\hat g}~ \vert J, r_A\ra
\label{wigner}
\eeq
where $l_A,~r_A$ label the states within the representation $J$.

We now define the right and left translation operators on $g$ as 
\beq
R_A ~g = g~T_A~~~~~~~~~~~~~~~~~~~~L_A ~g = T_A~g
\label{RL}
\eeq
where $T_A$ are the $SU(k+1)$ generators in the representation to which $g$ belongs. 
The $U(1)$ gauge field in (\ref{NA}) changes by a gauge transformation under a right $U(1)$ rotation of the form $g \rightarrow gh$,  while it remains invariant under an $SU(k)$ right rotation. This implies that in the case where the fermions couple only to the abelian gauge field $a$, the corresponding single particle wavefunctions have a fixed $U(1)_R$ charge and are singlets under $SU(k)$ right rotations \cite{KN2}. In particular the wavefunctions obey the condition
\beqar
 R_a ~\Psi_{m} &=& 0,~~~~~~~~~~~~~~~~a=1,\cdots,k^2-1  \nonumber \\
{R}_{k^2 +2k} ~\Psi_{m} &=& - {n k\over \sqrt{2 k
(k+1)}}~\Psi_{m} 
\label{9}
\eeqar
On the other hand, the nonabelian gauge field $\bar{A}^a$ in  (\ref{NA}) is invariant under right $U(1)$ rotations but noninvariant under right $SU(k)$ rotations. So in the case where the fermions have nonabelian degrees of freedom and couple to the full $U(k)$ background gauge field, the wavefunctions have the same fixed $U(1)_R$ charge as in (\ref{9}) but under right rotations transform as a particular $SU(k)$ representation $J'$ of dimension $N'= {\rm dim} J'$ \cite{KN3}. In this case,
\beqar
 R_a ~\Psi_{m; a'}  &=& 
\Psi_{m;b'} ~ (T_a)_{b' a'} \nonumber \\ 
{R}_{k^2 +2k} ~\Psi_{m; a'} &=& - {n k \over \sqrt{2 k
(k+1)}}~\Psi_{m; a'} 
\label{9NA}
\eeqar
The indices $a' ,b'~=1,\cdots, N'$ label the states within the $SU(k)$ representation $J'$ and can be thought of as the internal degrees of freedom of the nonabelian fermions coupled to the $U(k)$ background field. The matrices $T_a$ are the $SU(k)$ generators in the representation $J'$.

The coset operators $R_{\alpha}$ correspond to covariant derivatives while the $SU(k+1)$ operators $L_A$ correspond to magnetic translations. 
The Laplacian for the space is given by $-\nabla^2 = R_{+I}R_{-I} + R_{-I} R_{+I}
= 2 R_{+I}R_{-I} + {\rm constant}$. The minimum of the Hamiltonian, and hence the lowest Landau level, is
given by wavefunctions obeying 
\beq
R_{-I} ~\Psi_{m;a'} =0
\label{diff6}
\eeq
So, for the lowest Landau level in addition to the conditions (\ref{9NA}), 
$\vert J, r\ra$ must be a lowest weight state with $T_{-I} \vert J, r\ra =0$;
we will denote these states as $\vert a', -n\ra$.
Once the representation $J'$ is specified, one can identify representations $J$ of
$SU(k+1)$ which contain such a state. For example, if
there is no $SU(k)$ field, the symmetric rank $n$ representation of $SU(k+1)$
will contain the lowest weight state $\vert -n\ra$, which is an $SU(k)$ singlet.
The dimension of the $J$ representation which defines the dimensionality of the LLL Hilbert space is then given by $N= (n+k)!/n!k!$. 
In the nonabelian case where the fermions couple to the $U(k)$ background field LLL wavefunctions  form an irreducible $SU(k+1)$ representation $J$ whose lowest weight state is an $SU(k)$ representation $J'$. Since the $U(1)$ charge is fixed in terms of $n$, there are constraints on the type of allowed $J'$ $SU(k)$ representations \cite{KN3}. The dimension $N$ of the $SU(k+1)$ representation $J$ depends now on the particular $J'$ representation chosen, but for large $n$
\beq
N = {\rm dim} J \rightarrow {\rm dim} J' ~{n^k \over k!} = N' ~{n^k \over k!}
\label{35a}
\eeq

\section{Edge and bulk effective action}

We are now going to take the semi-clasical limit $N \gg K \gg1$ (large-$n$ limit) to extract the effective action $S_0$ describing the dynamics of the nonabelian LLL quantum Hall droplet. For this we need to derive expressions for the symbols of various operators and the star product. 
(The same expressions, but with $N'=1$, $\bar{A}=0$, apply for the abelian droplet.) Using the definitions (\ref{symb}, \ref{star}) and the LLL wavefunctions obtained earlier we find
\beqar
X*Y  =  XY + {1 \over n} P^{ij} D_i X D_j Y - {i \over n^2} P^{il} P^{kj} D_i X \bar{F}_{lk} D_j Y \nonumber \\
+  { 1 \over {2 n^2}} P^{ik} P^{jl} {\cal{D}}_iD_j X {\cal{D}}_kD_l Y + {\Or}({1 / n^3})
\label{47}
\eeqar
where $\bar{F}_{lk} = \bar{F}^a_{lk} (T_a)^T$,
$P^{ij} = g^{ij} + {i \over 2} (\Omega^{-1})^{ij}$ and  $D_i X  = \del_i X + i [\bar{A}_i ,~X]$. ${\cal{D}}_i$ is the Levi-Civita covariant derivative for a curved space 
such as ${\bf CP}^k$, namely, $
{\cal{D}}_iD_j X  \equiv  D_iD_j X - \Gamma^l_{ij} D_l X $,
where  $\Gamma_{ij}^l$ is the Christoffel symbol for ${\bf CP}^k$. $g_{ij}$ is the metric on ${\bf CP}^k$.

Equation (\ref{47}) is valid for both abelian and nonabelian cases. In the abelian case, $ X,~Y $ are commuting functions and $\bar{F}_{lk} \rightarrow 0$, so that $D_i X \rightarrow \del_iX$ and ${\cal{D}}_iD_j X \rightarrow  \del_i\del_j X -\Gamma_{ij}^l \del_l X$.

The derivation for the action is simplified if we use an $SU(k)$ invariant confining potential $V$, so that the ground state has spherical symmetry \cite{KN3}. In this case we find that the symbol corresponding to the density matrix is
\beq
(\rho_0)_{a'b'}   =  \rho_0(r^2) \delta_{a'b'} ,\quad
\rho_0(r^2) =  \Theta \Big(1 - {{R^2 r^2} \over {R_D^2}}\Big)
\label{step}
\eeq
where $\Theta$ is the step-function and $R_D$ is the radius of the droplet. In the semiclassical limit the density matrix is constant over the phase volume occupied by the droplet and its derivative is essentially a delta-function with support at the droplet boundary. As a result the effective action $S_0$ in the absence of fluctuating gauge fields is a purely boundary action. It can be written in terms of a matrix valued field $G= e^{i \Phi} \in U(N')$, where $\Phi$ is the symbol corresponding to $\hat{\Phi}$ in $\hat{U} = e^{i \hat{\Phi}}$, as 
\beqar
S_0  &=& {1 \over {4 \pi}} \tr \Biggl[ \int _{\del D}  ~
  \left[ \left( G^{\dagger} {\dot G} + \omega ~G^{\dagger} \L G \right)  
G^{\dagger} \L G  \right] \nonumber \\
&+& {1\over {4 \pi}} \int_{D} \tr \left[ -d \left( i \bar{A} dG G^{\dagger} + i \bar{A} G^{\dagger}dG \right) + {1 \over 3} \left( G^{\dagger}dG \right)^3 \right] \wedge \left( {n\Omega \over {2 \pi}}\right)^{k-1} {1 \over (k-1)!} 
\label{70}\\
&\equiv& S_{\rm WZW}(A^L=A^R=\bar{A}) \nonumber
\eeqar
where, $\L= {1 \over n} (\Omega^{-1})^{ij} \hat{r}_j D_i$ is the component of the covariant derivative $D$ perpendicular to the radial direction, along a special tangential direction on the droplet boundary;
$\hat{r}_i$ is the radial unit vector normal to the boundary of the droplet and 
$\omega = ({1 \over n} {\del V \over {\del r^2}} )_{\rm boundary}$ .

The action $S_0$ in (\ref{70}) is a higher dimensional generalization of a chiral,  Wess-Zumino-Witten action, vectorially gauged
with respect to the time-independent background gauge field $\bar{A}$. 
The third term is a WZW-term written as  an integral over a $(2k+1)$-dimensional region, corresponding to the droplet and time. The usual 3-form in the integrand of the WZW-term, $(G^\dagger d G)^3$, has now been augmented to the appropriate $(2k+1)$-form $(G^\dagger d G)^3 \wedge (\Omega)^{k-1}$. Since the WZW-term is the integral of a locally exact form, the whole action $S_0$ is a boundary action.

In the presence of gauge interactions the action describing the LLL dynamics is given by (\ref{10}). A straightforward but rather tedious calculation \cite{KA, review},  following the idea outlined after eq.(8) determines $\A$ in terms of the gauge fields $A_{\mu}$ (up to additional gauge invariant terms),
\beqar
\A =  A_0 &- & {i \over 2n} g^{ij} \left[ A_i,~2 D_i A_0 - \del_0 A_i + i [A_i,~A_0] \right] 
+ { 1 \over {4n}} (\Omega^{-1})^{ij} \{A_i, 2 D_j A_0 - \del_0 A_j + i [A_j,~A_0] \} \nonumber \\
&+& u^i A_i -{i \over {2n}} g^{ij} \left[A_i,~A_k \right] \del_j u^k + {1 \over {4n}} (\omega_K^{-1})^{ij} \{A_i,~A_k \} \del_j u^k \nonumber \\
&- &{i \over {2n}} g^{ij} \left[ A_i,~  2 D_j A_k - D_k A_j + i [ A_j,~A_k] ~ + 2 \bar{F}_{jk} ~\right] u^k \nonumber \\
&+ &{1 \over {4n}} (\Omega^{-1})^{ij} \{ A_i,~  2 D_j A_k - D_k A_j + i [ A_j,~A_k] + 2 \bar{F}_{jk} ~ \} u^k \nonumber \\
&+& {1 \over {2n^2}} g^{ik} (\Omega^{-1})^{jl} \left( {\cal{D}}_i A_j + {\cal{D}}_j A_i \right) \nabla_k\del_l V
\label{66}
\eeqar
where $[~,~]$ and $\{~,~\}$ indicate commutators  and anticommutators respectively and $u^i = { 1 \over n} (\Omega^{-1})^{ij} \del_j V$.
$u^i$ is essentially the phase space velocity, if we think of the LLL as the phase space of a lower dimensional system, with symplectic structure $n \Omega$ and Hamiltonian $V$.

Substituting (\ref{66}) into (\ref{10}) we find that the effective action $S$ splits into two pieces, a boundary action $S_{\rm edge}$ and a bulk one $S_{\rm bulk}$. 
The boundary action is essentially a gauged version of the higher-dimensional WZW action encountered in (\ref{70}). The gauging however appears in a left-right asymmetric way, indicating that the edge action by itself is not gauge-invariant. In particular,
\beq
S_{\rm edge} = S_{\rm WZW} (A^L, A^R)
\label{71}
\eeq
where $A^L = A+\bar{A}$, $A^{R} = \bar{A}$.

The bulk contribution to the action is given by
\beqar
\fl
S_{\rm bulk} & = & -{N \over {N'}} \int dt d\mu~ \rho_0~ \tr \left(A_0 + u^k A_k \right) \nonumber \\
& + & {kN \over {4 \pi n N'}} \int dt \rho_0\left[ \tr \biggl( (A+\bar{A}) d (A+ \bar{A}) + {2i \over 3} (A+\bar{A})^3 \biggr) \wedge \left( {\Omega \over 2\pi} \right)^{k-1} \right.\nonumber \\
& - &{(k-1) \over 2\pi }\left. \tr \biggl[ \biggl( (A+\bar{A})d(A+ \bar{A}) +{2i \over 3} (A+\bar{A})^3  \biggr) dV \biggr] \wedge \left( {\Omega \over 2\pi} \right)^{k-2}\right] \nonumber \\
& + & {N \over {2nN'}} \int dt d\mu ~\rho_0~ \tr \Big[    \nabla^i F_{ik}   + (k+1)  A_k
\Big] u^k
\label{77a}
\eeqar
If we consider the approximation where $R$ becomes large and the gradients of the external field are small compared to $B$, the metric-dependent terms in the last line of (\ref{77a}) can be neglected compared to the rest of the terms. In fact, with a little bit of algebra and using that $da= n\Omega$ and $N/N' = n^k/k!$ for large $n$, the 
``topological" part of (\ref{77a}) can be written as a single $(2k+1)$-dimensional Chern-Simons term
\beq
S_{\rm bulk} = S_{CS} (\tilde{A}), \qquad
\tilde{A} = \Big( A_0 +V,~ -a_i+\bar{A}_i+A_i \Big)\label{81c}
\eeq
in agreement with \cite{nair}.

The bulk action is not gauge-invariant. This has to do with the fact that a Chern-Simons action defined on a space with boundary is not gauge-invariant, the non-invariance given by a surface term. It is straightforward to check that the edge WZW action in (\ref{71}) exactly cancels the gauge-anomaly of the Chern-Simons term rendering the total effective action gauge-invariant as expected. This is a higher dimensional and nonabelian analog of the anomaly cancellation between the edge and bulk actions, well known in the case of two-dimensional planar QHE \cite{wen}.

\section{Summary}
An interesting observation which emerges from our analysis is that there is a universal one-dimensional matrix action describing the LLL dynamics, independent of the dimensionality and the abelian or nonabelian nature of the underlying fermionic system. The Hilbert space corresponding to the LLL states for the QHE on a space $\cal{M}$ defines a fuzzy version of $\cal{M}$. (In particular, the LLL of the example we have analyzed gives a definition of fuzzy ${\bf CP}^k$.) Using the star-product of this fuzzy $\cal{M}$ one derives the action of a noncommutative bosonic field theory, leading to an exact bosonization method at the level of the action. The semiclassical limit of this describes the dynamics of the corresponding $\nu=1$ quantum Hall droplet and it naturally separates into a boundary contribution and a bulk contribution which are interesting higher dimensional generalizations of the Wess-Zumino-Witten and Chern-Simons action respectively. These ideas can be easily extended to describe the dynamics of the $\nu=n$ quantum Hall droplet, where $n$ Landau levels are filled. 

In two dimensions the corresponding WZW and Chern-Simons actions define conformal field theories \cite{wen}. The common origin of their higher dimensional counterparts, via the matrix formulation, suggests the possibility that these theories may share similar features, such as conformal symmetry and chiral algebra. 

Although this presentation is in the context of QHE our analysis is relevant in the bosonization of a noninteracting fermionic system in higher dimensions by viewing it in phase space as a Landau problem with the symplectic structure being the magnetic field. Related work on phase space Hall droplets has been done in \cite{poly1}. Other approaches towards bosonization in higher dimensions have appeared in \cite{boso1}-\cite{boso7}.

A connection of our work to the Bergman kernel which can be used to approximate metrics on arbitrary K\"ahler manifolds is recently outlined in \cite{douglas}. The Bergman kernel is essentially the density matrix $\rho$ projected onto the LLL of the QHE defined on a manifold $\cal{M}$.
The Bergman metric is defined by $g_B= {1 \over n} \del \bdel \log \rho$ and provides, for large $n$, an approximation to Einstein metrics for K\"ahler manifolds embedded in ${\bf CP}^N$. In fact, an asymptotic expansion of $\rho$ in $1/n$ is in terms of curvatures of these spaces \cite{math}. In our case such an expansion can be obtained as the variation of the effective action $S$ with respect to $A_0$. Our formulation provides a physical context to obtain this in the case of both abelian and nonabelian external gauge fields and compare with existing results in mathematical literature \cite{math}.

\section{References}
\medskip


\end{document}